# On Side Channel Cryptanalysis and Sequential Decoding

Andreas Ibing

Technische Universität München
Arcisstrasse 21, 80333 Munich, Germany

***Abstract*—This paper presents an approach for side channel cryptanalysis with iterative approximate Bayesian inference, based on sequential decoding methods. Reliability information about subkey hypotheses is generated in the form of likelihoods, and sets of subkey hypothesis likelihoods are optimally combined into key bit log likelihood ratios. The redundancy of expanded keys in multi-round cryptographic schemes is exploited to correct round key estimation errors. This is achieved by sequential decoding, where subkey candidates are sorted by a probabilistic path metric and iteratively extended. The M-algorithm is presented as a concrete implementation example with deterministic runtime behaviour. The resulting algorithm contains previous hard decision differential analysis as special case for single-round analysis and M=1, and is strictly more accurate otherwise. The trade-off between estimation accuracy and complexity is scalable by parameter choice. The proposed algorithm is simulatively shown in an example scenario to reduce the number of required side channel traces compared to standard differential analysis by a factor of two when run with reasonable complexity, for the whole investigated signal-to-noise ratio range.**

## I. Introduction

Side channel analysis is used to infer secret keys of cryptographic systems from measurements of some physical processing leakage, especially power consumption and electromagnetic radiation. Analysis methods are categorized into 'simple' and differential methods. 'Simple' analysis directly classifies measured side channel traces, e.g. simple power analysis (SPA) and simple electromagnetic analysis (SEMA). Differential methods like differential power analysis (DPA) and differential electromagnetic analysis (DEMA) use a-priori knowledge about a subkey dependency of some intermediate processing result [1], [2], [3]. This a-priori knowledge is commonly described as differential analysis selection function, which describes the dependence of an information leaking variable on a few key and data bits. Data bits in this respect may be cyphertext bits when analyzing decryption or alternatively plain text bits when analyzing encryption. The standard method for differential analysis is to partition the traces according to a subkey hypothesis and the selection function, and to perform a difference-of-means test with a side channel leakage model. The most common model for power consumption and electromagnetic radiation (which is proportional to the derivative of power consumption) is the Hamming distance model, i.e. to assume differential side channel leakage proportional to the number of switching bits on a bus [4]. The number of subkey bits in the selection function must be small enough to allow for enumeration: all possible subkey hypotheses are tested in turn, and the most likely one is chosen as estimation result. A detailed description of the standard differential analysis algorithm and comparison with several variations is given e.g. in [5].

The main motivation of this paper is the following: many block cyphers like DES [6] and AES [7] are based on a multi-round scheme, where the key is expanded into the round keys according to a standardized key schedule. The key expansion can in fact be seen as a block code, with a code rate of $1/R$ for a scheme with $R$ rounds. This redundancy can in principle be exploited for soft-decision error correction in the side channel based key estimation. The optimality criterion is maximum likelihood sequence estimation (MLSE), i.e. to find or approximate the most likely valid expanded key. The idea faces two obstacles, for which this paper proposes solutions. First, the set of possible round keys is by design too large for an enumeration. The presented approach generates key bit log-likelihood-ratios (LLRs) from differential analysis, which allows to sort the key space according to reliability and to perform an informed search by only visiting a comparatively small number of candidate key sequences. Second, to follow the decryption (or encyption analogical) over several rounds, either the available cyphertext has to be conditionally decrypted after each round based on a round key sequence hypothesis, or equivalently the selection function could be adapted. This paper uses sequential decoding [8], [9], [10] to sort round key sequence hypotheses with a path metric, and to iteratively expand the most likely candidates.

Sequential decoding methods are common knowledge in communication theory and especially applied to multiple input multiple output (MIMO) demapping and decoding of concatenated channel codes (e.g. [11] and the references therein). The underlying problem is also to probabilistically and iteratively combine soft information from different sources. The contribution of this paper is therefore to apply these well-known methods from the area of communications to the area of side channel cryptanalysis. The result is a characterization of the three-dimensional trade-off between the number of required side channel traces, the measurement signal-to-noise ratio (SNR) and the applied computational effort.

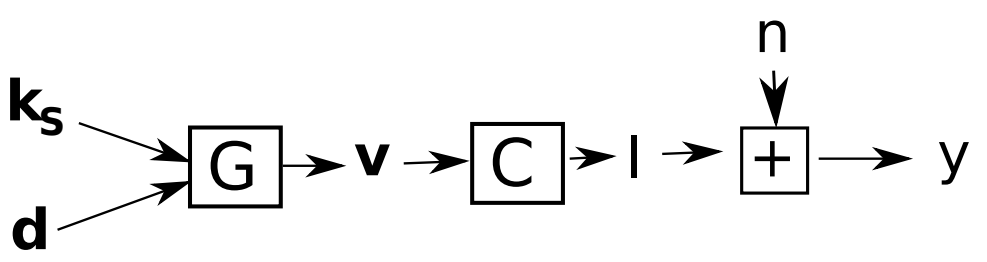

Fig. 1. Overview of variable dependencies.

The remainder of the paper is structured as follows. The next section describes the system model and notation, Sec. III shortly reviews standard single-bit differential analysis. The proposed algorithm is presented in detail in Sec. IV including key bit LLR generation, sequential decoding and multi-round soft information combining. The complexity of this algorithm in dependence on parameter choices is evaluated in Sec. V. The accuracy of the estimation is simulatively evaluated and compared to the standard hard-decision differential analysis in Sec. VI for different SNR values. The last section discusses the results and possible modifications of the presented approach.

## II. System Model and Notation

Vectors are denoted in bold font. The expectation operator is denoted $\mathbb{E}[\cdot]$, variance as $\mathbb{V}[\cdot]$ and probability as $\mathbb{P}(\cdot)$. A multi-round cryptographic scheme with $R$ rounds is assumed. The expanded key is written as

$$\mathbf{k} = (\mathbf{k_1} \dots \ \mathbf{k}_R) \ , \tag{1}$$

with the round keys $\mathbf{k}_i, \ i = 1 \dots R$. Each round key consists of $N_B$ bits.

### *A. Sensitive variables and selection function for differential analysis*

Differential analysis requires a-priori knowledge about at least one sensitive intermediate variable, which leaks information through the side channel [1]. A typical example are byte-wise S-box lookups for substitution in a substitution-permutation network like AES [4]. The selection function depends on the specific cryptographic algorithm, and also on implementation aspects like e.g. the usage of masking [12]. The sensitive variable is denoted $\mathbf{v} = (v_1 \dots v_{N_V})$, and the selection function $G$:

$$\mathbf{v} = G(\mathbf{d}, \mathbf{k_S}) \tag{2}$$

$\mathbf{v}$ depends on the subkey $\mathbf{k_S}$ which comprises several bits of the round key $\mathbf{k_i}$, and on some data bits $\mathbf{d}$. The length (number of bits) of the subkey is left variable at this point.

### *B. Power consumption model and feature extraction*

The power consumption of digital devices depends on their transistor switching activity. The two most common power consumption models for side channel analysis are the Hamming weight model and the Hamming distance model [4]. Their applicability depends on the implementation technology of the device under test.

*1) Hamming weight model:* This model assumes a differential current proportional to the number of positive bits on a bus (i.e. the Hamming weight). It is applicable for devices implemented in precharge logic [4]. The differential side channel leakage $l(t)$ of the sensitive variable under this model is:

$$l(t) = \sum_{i=1}^{N_V} v_i(t) \tag{3}$$

*2) Hamming distance model:* The Hamming distance model assumes that the switching of bits on a bus consumes an amount of current proportional to the number of switching bits (i.e. the Hamming distance of consecutive words on the bus). This model is applicable for devices implemented in CMOS logic. The side channel leakage of the sensitive variable (e.g. at transfer to/from a register) thus depends on the previous word on the bus.

*a) Device using software implementation::* If the cryptographic algorithm is implemented in software, the assumption can be made that the sensitive variable is handled at a fixed point in the program. Then a constant but unknown previous bus state $\mathbf{r}$ can be assumed [13]. The differential leakage becomes:

$$l(t) = \sum_{i=1}^{N_V} \Big( v_i(t) \oplus r_i \Big) \tag{4}$$

with $\oplus$ denoting the XOR operator. [13] proposes to assume all possible reference states $\mathbf{r}$ and to use the most likely one, computed as highest correlation factor.

*b) Device using hardware implementation::* If the cryptographic algorithm is implemented in hardware, it might be assumed that consecutive values of the sensitive variable are stored consecutively in the same register. The differential leakage then becomes:

$$l(t) = \sum_{i=1}^{N_V} \Big( v_i(t) \oplus v_i(t-1) \Big) \tag{5}$$

*c) Comparison::* a generalized power model comprising both the Hamming weight and the Hamming distance model with arbitrary parameters is obtained by conceptually assuming a convolutional code $C$ over the sensitive variable on the word level (with 0 or 1 tap delay in the presented examples):

$$l(t) = \sum_{i=1}^{N_V} C(\mathbf{v}(t), \mathbf{v}(t-1)) \tag{6}$$

With respect to the key bits $\mathbf{k}_S$, the resulting differential leakage

$$l(t) = \sum_{i=1}^{N_V} C\Big(G(\mathbf{d}(t), \mathbf{k_S}(t)), G(\mathbf{d}(t-1), \mathbf{k_S}(t-1))\Big) \tag{7}$$

can be seen as a short block code concatenated with a short convolutional code (for decoding of concatenated codes see e.g. [14], [15], [10]). Unknown code parameters need to be known a-priori, guessed or estimated.

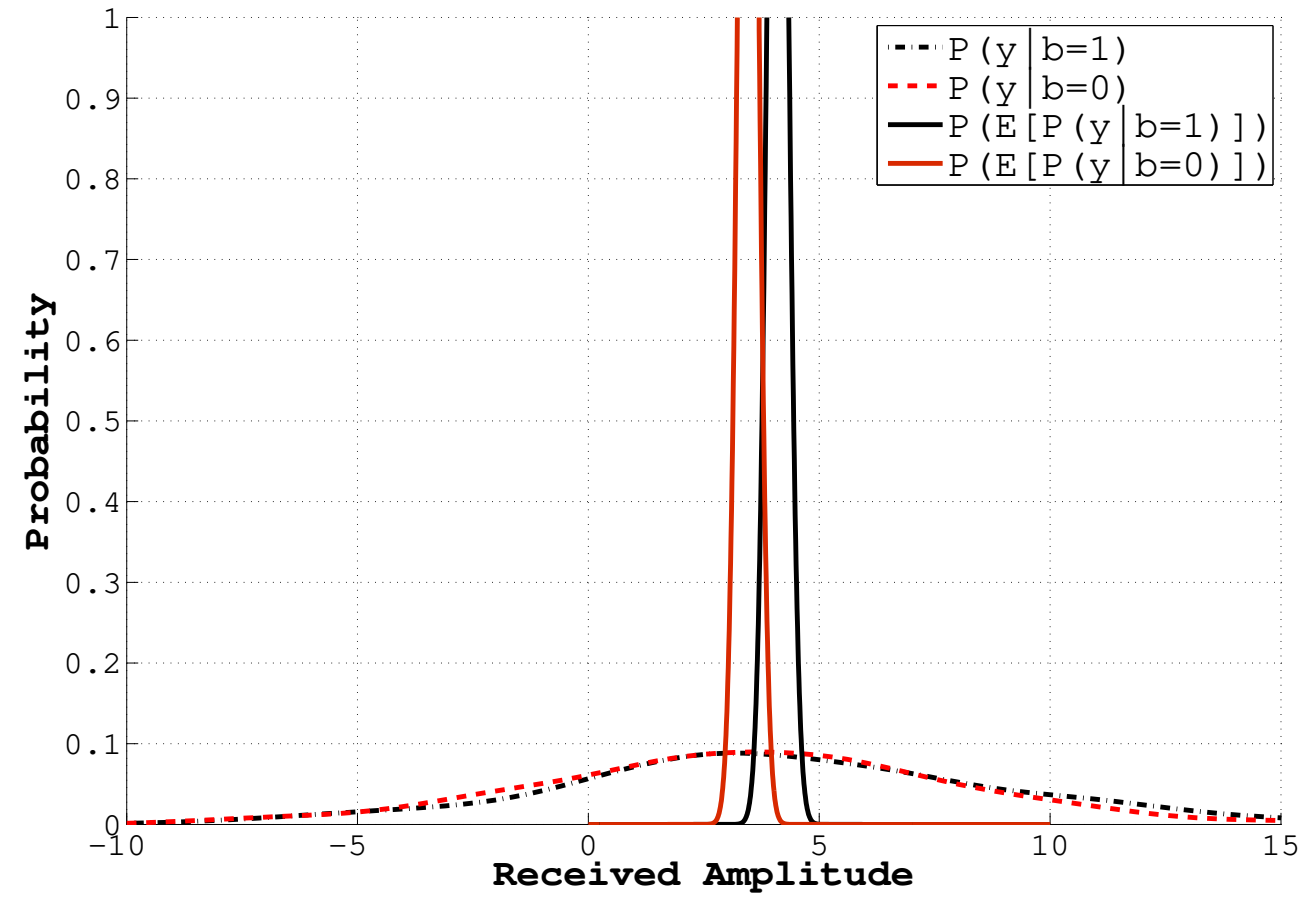

Fig. 2. Differential analysis: conditional densities of received amplitude and mean estimation functions for 500 traces at SNR=6dB.

*3) Feature extraction:* Side channel measurements usually use oversampling with respect to the clock of the device under test, especially when measuring an electromagnetic side channel. Measurements $y$ use the clock $\tau$, as compared to the device using clock $t$. Since the device's switching power is of interest, and because power is proportional to the squared amplitude of voltage or current (or the related fields respectively), the received energy during one target clock cycle may be extracted as feature:

$$f(t) = |y(t)|^2 = \sum_{\tau=1}^{N_S} |y(t,\tau)|^2 \ , \tag{8}$$

where $N_S$ is chosen to capture the clock's rising edge half period where the information transfer happens.

*4) Assumption for simulations in this paper:* For the sake of clarity of the presentation, with the focus lying on key bit LLR generation and multi-round soft information combining, the Hamming weight model, Eq. (3), is assumed. As discussed, a Hamming distance model could be reduced to the Hamming weight model by assuming a serially concatenated short convolutional code over the sensitive variable. Additive noise $n(t)$ on the measurements caused by the measurement equipment itself (like thermal noise of the receive amplifier) as well as environmental noise and interference from other switching activity in the circuit is assumed independent from the sensitive variable. Simulations in this paper assume therefore:

$$y(t) = l(t) + n(t) \ , \tag{9}$$

with the SNR:

$$\gamma = \frac{\mathbb{V}[l]}{\mathbb{V}[n]} = \frac{\sigma_l^2}{\sigma_n^2} \tag{10}$$

A diagram of the variable dependencies is shown in Fig. 1.

## III. Review of Standard Single-Bit Differential Analysis

The power dissipation of indidivual switching bits is too small to be recognizable with 'simple' analysis. Fig. 2 shows the (simulated) distribution of received amplitudes, conditioned on one bit $v_i$ being 1 or 0 respectively, i.e. $\mathbb{P}(y|v_i = 1)$ and $\mathbb{P}(y|v_i = 0)$. The figure shows almost identical conditional distributions at 6dB SNR.

As 'standard' single-bit differential analysis it is referred to [1], [2], [3]. With the known selection function and a subkey hypothesis $\mathbf{k}_S$, the trace data can be partitioned according to the value of one $v_i$. For this subkey hypothesis, the difference of means over the disjoint trace set partitions is computed:

$$P_D = \mathbb{E}_T[f|v_i = 1] - \mathbb{E}_T[f|v_i = -1] \tag{11}$$

For an incorrect subkey hypothesis, the expected result is zero. For a correct subkey hypothesis, a nonzero value corresponding to the bit's switching energy is expected. Independent noise is averaged away by the expectancy over the trace partitions. The limit for an increasing number of traces $N_T$ is:

$$\lim_{N_T \to \infty} P_D = \begin{cases} \epsilon & \text{if } \mathbf{k}_S \text{ correct,} \\ 0 & \text{else} \end{cases} \tag{12}$$

with $\epsilon$ being the received switching energy. The subkey is estimated by testing all hypotheses and choosing the one with highest difference of means, corresponding to the maximum likelihood (ML) subkey.

## IV. Proposed Algorithm

The proposed algorithm is an approximative maximum likelihood sequence estimation (MLSE), where the accuracy can be scaled by the invested computational effort. A difference to the standard differential analysis algorithm is that not only the most likely subkey hypothesis is used, which would mean a premature quantization of an intermediate result, but that all tested hypotheses are used and optimally combined using a probabilistic formulation. This probabilistic formulation also allows for combination of different selection function. For stochastic inference, key bits are modelled as independent random variables. Their probability distribution functions are computed from partial problems, and iteratively updated according to Bayes' theorem whenever an additional problem constraint or information source is joined (Bayesian network, Belief propagation [16]). Quantization of key bit probabilities into binary values is done only at the end of the algorithm, to avoid error propagation. Since the key search space even in one round is by design for too large to consider all possibilities, an informed graph search is used which avoids looking at most possible key values. This is enabled by iterative sorting of likelihoods of parts the joint solution and pruning the search tree.

### *A. Problem structure: factorize expanded key probabiltiy density function*

The joint probability density function of all random variables is factorized using conditional probabilities and Bayes' theorem, to break down the estimation problem into smaller components, which can be treated with limited computational effort. This complexity reduction compared to joint estimation exploits conditional independencies between variables [16].

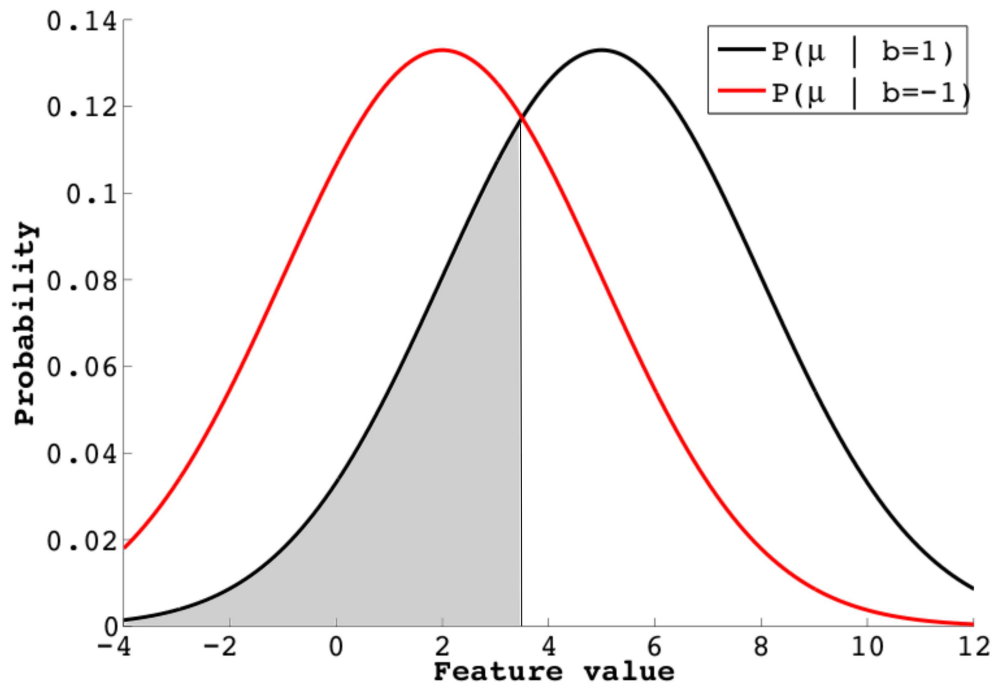

Fig. 3. Subkey likelihood generation from estimation function of means: Gaussian tail probability (shaded).

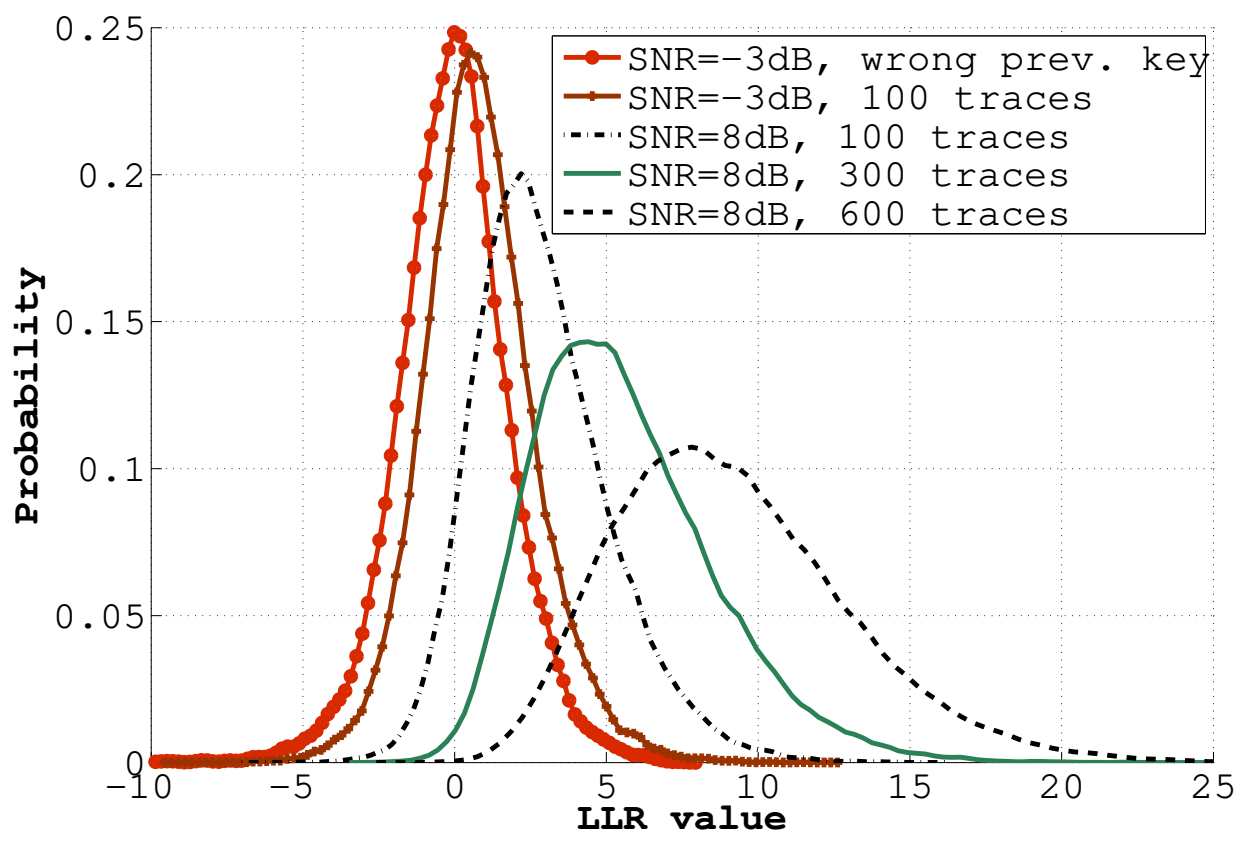

Fig. 4. LLR distribution for correct subkey hypothesis of 1 bit length and positive bit value, for different SNR and different number of traces. Distribution for a wrong previous round key estimate also shown.

For denotational as well as for computational convenience, bit values are denoted as $\{-1, +1\}$ in the following, instead of the normal $\{0, 1\}$. Factorizing the expanded key estimation problem into the round key estimation problems is:

$$\underbrace{\mathbb{P}(\mathbf{k}_{i\ldots1})}_{\text{posterior probability}} = \underbrace{\mathbb{P}(\mathbf{k}_i|\mathbf{k}_{i-1}\ldots\mathbf{k}_1)}_{\text{likelihood}} \cdot \underbrace{\mathbb{P}(\mathbf{k}_{i-1\ldots1})}_{\text{prior probability}} \quad (13)$$

where $\mathbf{k}_{i\ldots1}$ is the sequence of round keys $\mathbf{k}_i$ downto $\mathbf{k}_1$. Bayesian updating can then be applied iteratively over the rounds. The formulation with conditional probabilities turns the problem into a tree code model, where a graph search can be run for the 'correct', i.e. the most likely leaf. The sequential decoding approach offers several implementation algorithms, among them the stack algorithm [8], the M-algorithm [17] and the T-algorithm [18]. The stack algorithm corresponds to a depth first search, while the M-algorithm and T-algorithm correspond to a breadth first search, where the number of extended 'most likely' nodes per level is limited (constant for the M-algorithm and variable for the T-algorithm).

### *B. Generate soft information: key bit log-likelihood ratios*

Log-likelihood ratios (LLRs) are a convenient way to express a key bit's distribution function in one number. Another advantage of computations in the log domain is that multiplications necessary to compute the probability of a sequence of independent bits are effectively turned into additions. The LLR of a bit $b$ is:

$$L(b) = \log \frac{\mathbb{P}(b = 1)}{\mathbb{P}(b = -1)} \quad (14)$$

and the inverse:

$$\mathbb{P}(b = \pm 1) = \frac{e^{\pm L(b)/2}}{e^{+L(b)/2} + e^{-L(b)/2}} \quad (15)$$

The sign of the LLR indicates the likely bit value, and the LLR magnitude its reliability.

*1) Estimate subkey likelihoods:* To generate LLRs from differential analysis, the reproduction property of the Gaussian distribution is used: the estimation function of the mean value of a Gaussian distribution with $N_T$ samples is again Gaussian distributed, but with $N_T$ times smaller variance: Fig. 2 shows not only the conditional received values, but also the estimation functions of the conditional mean values. If it is:

$$\mathbb{P}(f|v_i = +1) \propto \mathcal{N}(\epsilon, \sigma_n) \quad (16)$$

then the estimation function of the mean $\mu_p$ for positive $v_i$ is:

$$\mathbb{P}(\mu_p) \propto \mathcal{N}(\epsilon, \frac{\sigma_n}{\sqrt{N_T/2}}) \quad (17)$$

The likelihood of the mean value $\mu_p$ for positive $v_i$ being larger than the mean value $\mu_n$ for negative $v_i$ can be approximated as the Gaussian tail distribution:

$$\mathbb{P}(\mu_p > \mu_z) \approx Q(\frac{(\hat{\mu_p} - \hat{\mu_z})/2}{\hat{\sigma}_\mu}) \quad (18)$$

where $\hat{\mu}_p$ and $\hat{\mu}_z$ are the conditional mean values computed over the trace partitions, and $\hat{\sigma}_\mu = \frac{\sigma_n}{\sqrt{N_T/2}}$ is the correspondingly computed standard deviation of the estimation functions (assuming equal partition sizes). The tail probability is illustrated in Fig. 3. Computation of this value can be implemented as a table lookup of the Q-function. The likelihood of the subkey hypothesis to be correct is the likelihood of $\mu_p$ being larger than $\mu_z$:

$$\mathbb{P}(\mathbf{k}_S) = \mathbb{P}(\mu_p > \mu_z) \quad (19)$$

Thus, likelihoods can be generated for all possible subkey hypotheses $\mathbf{k_S}$.

*2) Compute key bit LLRs from subkey likelihoods:* The number and length of the subkey hypotheses depend on selection function. For $s$ bits length, there are $2^s$ hypotheses. Each of the $i = 1 \ldots s$ bit positions is contained with positive bit value and negative bit value in $2^{s-1}$ hypotheses each. This entails that there is always at least one hypothesis with positive key bit value and one counterhypothesis with this bit being negative. The key bit LLRs can be computed as:

$$L(b_i) = \log \frac{\sum_{\mathbf{k}_S \in \mathcal{S}_{i+}} \mathbb{P}(\mathbf{k}_S)}{\sum_{\mathbf{k}_S \in \mathcal{S}_{i-}} \mathbb{P}(\mathbf{k}_S)} \quad (20)$$

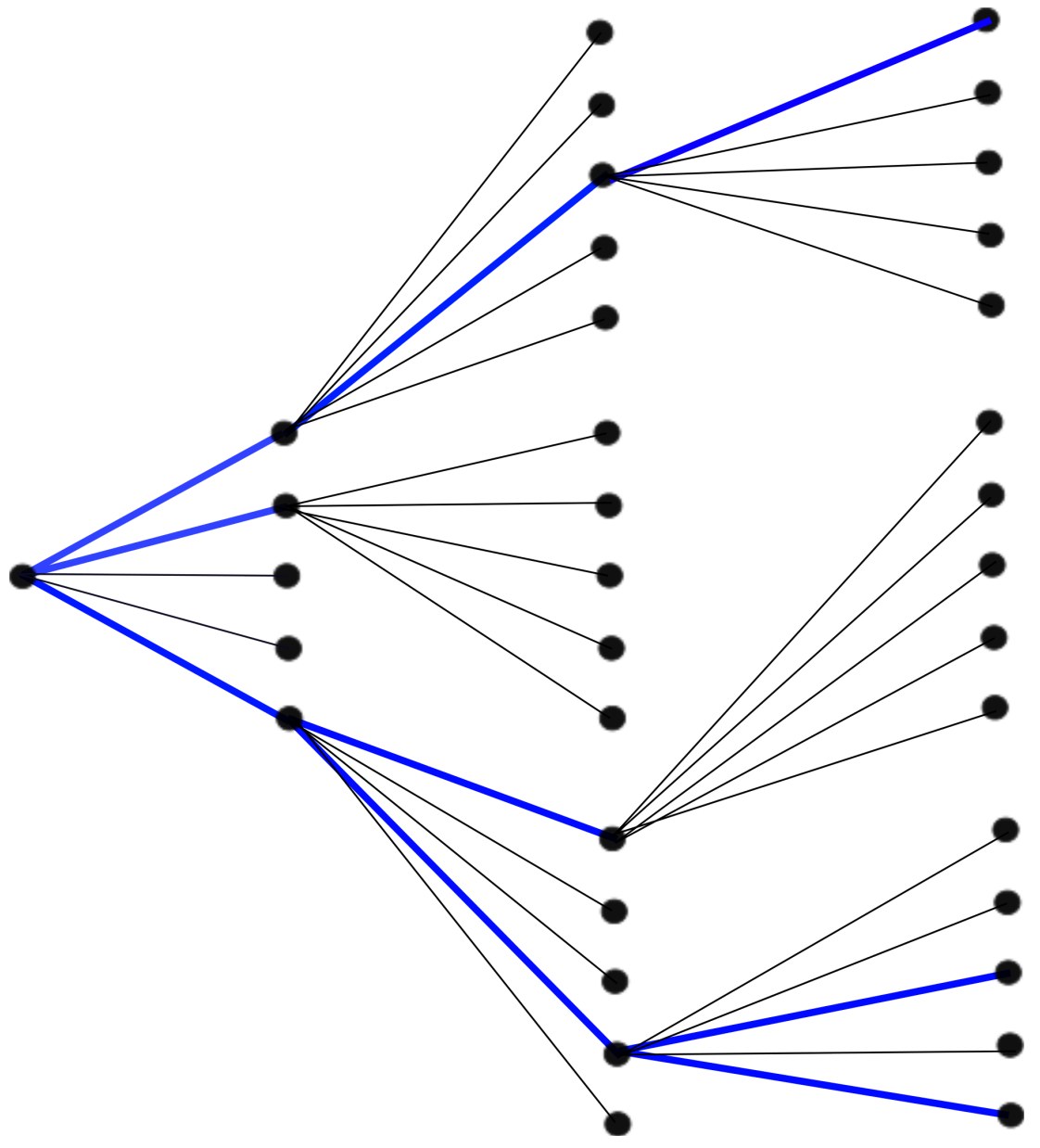

Fig. 5. M-algorithm: nodes are expanded into $M_E$ children, of which the most likely $M_R$ per level are further evaluated (example $M_E = 5,\ M_R = 3$).

where $\mathcal{S}_{i+}$ denotes the set of all hypotheses where bit number $i$ has positive value, and $\mathcal{S}_{i-}$ the set where it has negative value.

To reduce the computational effort, the Max-Log approximation [19] can be used:

$$\log \sum_i a_i \approx \max\Big(\ \log(a_i)\ \Big) \tag{21}$$

which yields

$$L(b_i) \approx \max_{\mathbf{k}_S \in \mathcal{S}_{i+}} \Big(\log\ \mathbb{P}(\mathbf{k}_S)\Big) - \max_{\mathbf{k}_S \in \mathcal{S}_{i-}} \Big(\log\ \mathbb{P}(\mathbf{k}_S)\Big) \tag{22}$$

Fig. 4 shows conditional LLR distributions for positive bit value, for different SNR and different number of traces. It also shows an LLR distribution for a wrong previous round key. The improvement in expected LLR magnitude and classification accuracy with increasing SNR and increasing number of traces is clearly visible.

*3) Combining independent selection functions:* If LLRs for the same bit are generated from independent selection functions, these LLRs are simply added for information combining. Different subkey hypothesis lengths are no problem in this formulation.

### C. Sequential decoding: multi-round soft information combining

The proposed implementation uses the M-algorithm, which is illustrated in Fig. 5. It is a breadth-first tree search with deterministic limited complexity. Each node in a tree is reachable from the root with exactly one path. Visited nodes are assigned the path metric, and in each tree level only the M-best nodes (best path metrics) are expanded into the next level. The algorithm finds the leaf with the 'approximately best' metric, where the approximation is better for larger M. The M-algorithm comprises the 'greedy' heuristic for the special case M=1.

*1) Path metric:* In order to find the MLSE solution for the expanded key, the *key bit sequence log-likelihood* is chosen as path metric. This is a computationally convenient choice because the Max-Log approximation can also be applied to the mapping from LLR to probability (Eq. (15)), e.g. [11]:

$$\begin{aligned} \log(\mathbb{P}(b_i = -1)) &= \log \frac{1}{1 + e^{L(b_i)}} \\ &= \log 1 - \log(e^0 + e^{L(b_i)}) \\ &\approx 0 - \max(L(b_i); 0) \end{aligned} \tag{23}$$

The Max-Log approximated log-likelihood of a sequence of bits, e.g. a round key, then becomes:

$$\log(\mathbb{P}(\mathbf{b})) \approx \sum_i L(b_i) \cdot b_i \tag{24}$$

Summands increase the sequence log-likelihood if both the L-value and the bit have equal signs; otherwise they reduce it. Bit positions contribute to the metric according to their reliability (LLR magnitude). An overview of the proposed algorithm in pseudo-code formulation is given in Alg. 1.

*2) Node expansion: generate child round key candidate set:* This section describes the node expansion into its child nodes. The child round key candidate set is generated from the LLRs for the current round's bit positions (which are obtained according to Sec. IV-B). Since the round LLR generation needs the data (cyphertext or plaintext) corresponding to the traces, this can be directly applied only in the first round. For later round $i$, the data first needs to be decrypted according to the current candidate's previous round keys $\mathbf{k}_{i-1\ldots1}$ (conditional decryption, Alg. 1). The children's round metrics $\beta_R$ therefore follow from Eq. (24) to:

$$\beta_R(\mathbf{k}_i|\mathbf{k}_{i-1\ldots1}) = \sum_{j=1}^{N_B} L(k_{i,j}|\mathbf{k}_{i-1\ldots1})\ \cdot\ k_{i,j} \tag{25}$$

By design of the cryptographic algorithm, it is again unfeasible to visit all child nodes even for a single node expansion. The expansion into the $M_e$ 'best' child nodes therefore again uses a heuristic.

One possible heuristic to find the '$M_E$-best' children of a node is described in the following. It computes round metrics $\beta_R$ for a search space of $M_S > M_E$ candidate childs which is determined as:

- sort conditional round LLRs $L(k_{i,j}|\mathbf{k}_{i-1\ldots1})$ according to magnitude
- the most likely child is determined as the LLR signs
- the other $M_S - 1$ child candidates are determined as bit deviations from the most likely child:
  - enumeration subspace: for the $n_E$ positions with smallest LLR magnitude (most unreliable), all $2^{n_E}$ bit combinations are enumerated
  - combinatorial subspace: for the $N_B - n_E$ remaining bit positions, up to $n_C$ bit deviations are considered

– the search space is the Cartesian product of the enumeration subspace and the combinatorial subspace

- the $M_S$ metrics are computed
- the $M_E$ childs with best metric are returned.

A more detailed description of this node expansion heuristic is listed as pseudo-code (similar to Octave/Matlab notation) in Alg. 2.

*3) Reduce candidate set with path metric:* The number of candidates on one tree level is then reduced to $M_R < M_E$ using the path metric $\beta_P$:

$$\beta_P(\mathbf{k}_{i..1}) = \underbrace{\beta_R(\mathbf{k}_i|\mathbf{k}_{i-1...1})}_{\text{round log-likelihood}} + \underbrace{\beta_P(\mathbf{k}_{i-1..1})}_{\text{a-priori log-probability}} \tag{26}$$

The comment in the underbraces notes that the extension of the path metric by one level is Bayesian updating (Eq. (13)) in the log domain when adding the new round's information, i.e. the a-posteriori log-probability is computed.

The reduction of the candidate set to $M_R$ candidates is possible because the expected round LLRs are zero for any wrong round key in the path candidate (compare Fig. 4). The expected best child's round metric is:

$$\mathbb{E}\Big[\max_{\mathbf{k}_i} \beta_R(\mathbf{k}_i|\mathbf{k}_{i-1...1})\Big] = \begin{cases} N_B * \mathbb{E}\Big[\big|L(k_{i,j}|\mathbf{k}_{i-1...1})\big|\Big] \\ \quad > 0 \text{ if } \mathbf{k}_{i-1...1} \text{ correct,} \\ 0 \text{ else} \end{cases} \tag{27}$$

which gives a significant difference in the path metric values for candidates which contain a false round key.

**input** : $T$ traces and data
**output**: 'most likely' expanded key candidate

```
// special treatment for round 1:
foreach Byte per round do
    LLRs ← getLLRs (traces,data) ;
end
// get M_E best round key candidates:
candidate [] ← getBestChilds (LLRs, M_E) ;
// other rounds:
for round r ← 2 to N_R do
    foreach candidate do
        rounddata ← roundDecrypt (data, candidate);
        foreach Byte per round do
            LLRs = getLLRs (traces, rounddata) ;
        end
    end
    /* compute path metrics, reduce to M_R candidates;  */
    candidate [] ← reduceCandidates (candidate [], M_R) ;
    foreach candidate do
        candidate [] ← getBestChilds (LLRs, M_E)
    end
end
return reduceCandidates (candidate [], 1) ;
```

**Algorithm 1:** Pseudo-code overview of implementation using M-algorithm, assuming byte-wise processing.

## V. Complexity

The complexity measure of interest for application would be the time complexity (number of clock cycles) on the hardware available to run the algorithm. But this cycle count depends on the instruction set architecture of the processor cores used, and special hardware acceleration like with FPGAs is possible. Therefore instead of assuming any special instruction set, a high level complexity evaluation in dependence on the parameters $M_E$, $M_R$, $n_E$, $n_C$, and $R$ is given (assuming the node expansion heuristic from Alg. 2).

*a) Number of visited nodes::* one child expansion visits $M_S$ candidates, with:

$$\begin{aligned} M_S &= 2^{n_E} \cdot \sum_{i=0}^{n_C} \binom{N_B - n_E}{i} \\ &= 2^{n_E} \cdot \sum_{i=0}^{n_C} \frac{(N_B - n_E)!}{(N_B - n_E - i)! \cdot i!} \end{aligned} \tag{28}$$

$M_E$ candidates are returned for each expansion, and the tree is pruned for each level to $M_R$ round survivors. The number of visited nodes is therefore:

$$N_{\text{visits}} = M_S + M_R \cdot M_S \cdot (R-1) \tag{29}$$

*b) Metric computations::* in a straight-forward implementation, one complete metric would be computed per visited node. But since the metric is additive over path segments (Eq. (26)), the reuse of intermediate results is possible. And since childrens' round metrics are computed as sums of the same LLRs with different signs (Eq. (25)), further reuse is possible with a Gray code binary enumeration [21].

*c) Compare/Select operations for candidate selection::* Child expansion needs to find the $M_E$ best values out of $M_S$. Round candidate reduction needs to find $M_R$ best out of $M_E$ after the first round, and out of $M_E \cdot M_R$ later. Different search algorithms like bubblesort or comb sort are possible, but other algorithms like insertion sort may be preferable for parallelization speedup on modern processors with SIMD instructions. Again, a reuse may be possible with adequate data structures.

*d) Number of round decryptions::* after each round, $M_R$ decryptions are needed, in sum:

$$N_{\text{decrypt}} = M_R \cdot (R-1) \tag{30}$$

## VI. Accuracy

A simulative evaluation of the estimation accuracy is shown in Fig. 6. The scenario assumes $R = 10$ rounds with $N_B = 128$ bit round key length, and a selection function

```
function Mbest ← getBestChildsApprox (LLRs,
M_E, searchBitsEnum, numCorrectErrors) ;

input  : round LLR vector conditioned on previous round
         key path hypothesis
output: M_E 'most likely' round key candidates

numL ← length (LLRs) ;
/* sort according to reliability:                    */
[Lsorted, Lperm ] ← sort (LLRs, 'descend') ;
bestCandidatePerm ← sign (LLRs) ;
sizeEnum ← 2^searchBitsEnum ;
searchEnum ← de2bi(0:sizeEnum-1, 'left-msb') ;
/* bit values ±1:                                    */
zIdxs ← find (searchEnum == 0) ;
searchEnum (zIdxs) ← −1 ;
candCombIdxs ← cell(numCorrectErrors,1) ;
nCombSep (0) ← 1 ;
nComb ← nCombSep (0) ;
for i=1:numCorrectErrors do
    candCombIdxs(i) ← nchoosek (1:128 -
    searchBitsEnum,i) ;
    nCombSep (i) ← nchoosek (128 -
    searchBitsEnum,i) ;
    nComb ← nComb + nCombSep (i) ;
end
numCand ← nComb * 2^searchBitsEnum ;
candCombBinary ← ones (numCand, numL) ;
pos ← 1 ;
for i ← 1:numCorrectErrors do
    for j ← 1 : nCombSep (i-1) * sizeEnum do
        pos ← pos + 1 ;
        /* deviations from most likely
           round candidate:                          */
        candCombBinary (pos, candCombIdxs(i)(mod
        (j,nCombSep (i))+1,:) ) ← -1 ;
        candCombBinary (pos, numL-
        searchBitsEnum + 1 : end) ← searchEnum
        (mod (j,sizeEnum)+1,:) ;
    end
end
rCandPerm ← bestCandidatePerm.*
candCombBinary;
/* compute metrics, get M_E best:                    */
metrics ← ∑((rCandPerm.* kron (ones
(numCand,1),LSorted)).') ;
[candMetr, candIdx ] ← sort (metrics,'descend') ;
bestCandIdxsPerm ← candIdx (1:M_E) ;
bestRCandBinPerm ← rCandPerm
(bestCandIdxsPerm,:) ;
/* unpermutate initial sorting:                      */
[dummy, unpermIdxs ] ← sort (Lperm, 'ascend') ;
return Mbest ← bestRCandBinPerm (:,unpermIdxs) ;
```

**Algorithm 2:** Pseudo-code (similar to Octave / Matlab notation [20]) to approximately find best $M_E$ child candidates.

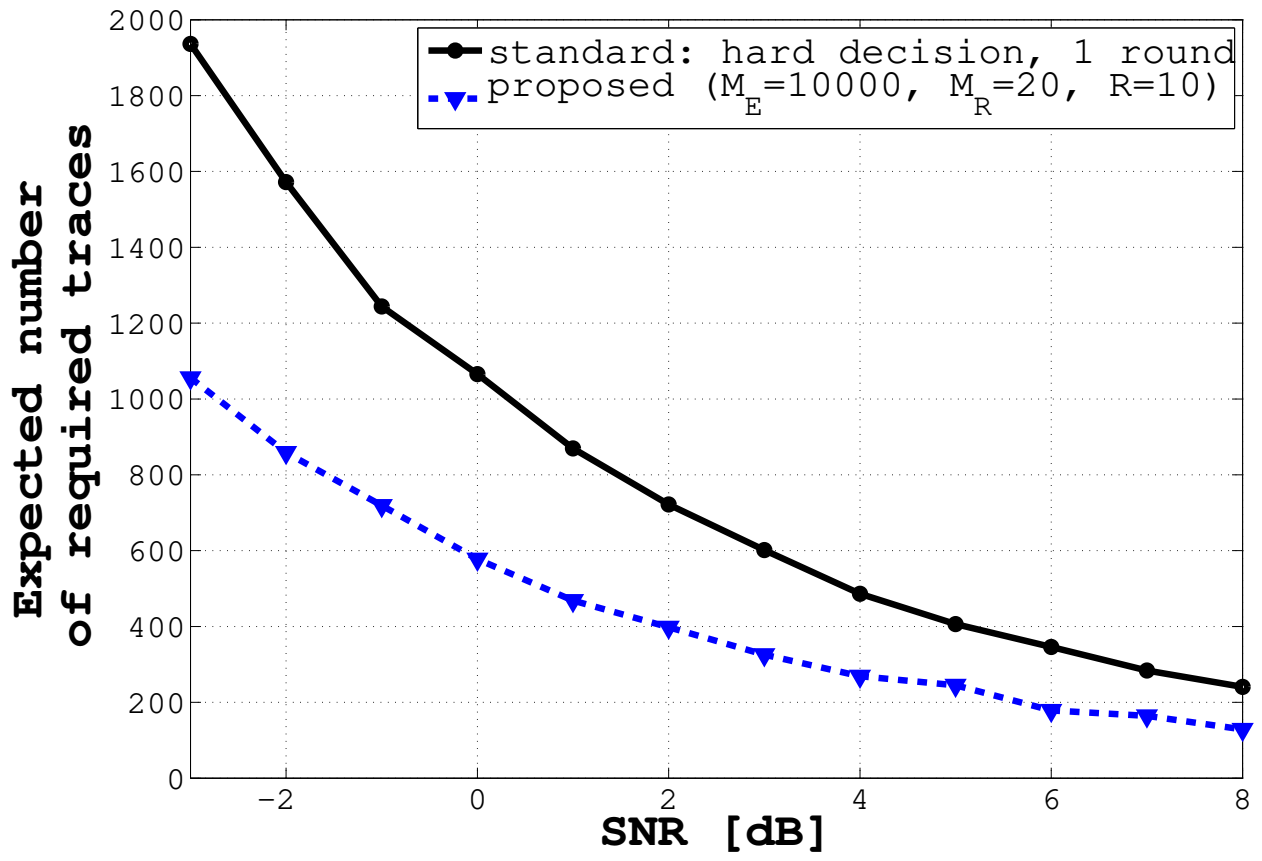

Fig. 6. The proposed algorithm reduces the number of required traces compared to the standard algorithm by a factor of two in a configuration with reasonable complexity.

with subkey length of 1 bit. The proposed algorithm is run with $M_E = 10000$ and $M_R = 20$ and compared for different SNR with the standard single-round 'hard decision' estimation (which directly quantizes to the ML subkey hypotheses). Simulation computes the expected number of required traces to successfully estimate the secret key. The simulation results are obtained with a scripting language on a single PC, which may be seen as indication of a reasonable computation complexity. Over the complete investigated SNR range, the proposed algorithm reduces the required number of traces roughly by a factor of two (corresponding to a 3dB SNR gain). Accuracy can be further improved by increasing $M_E$ and $M_R$.

## VII. Discussion

This paper describes approximative Bayesian inference for side channel secret key estimation in the sense of MLSE. The probabilistic formulation allows for optimal combination of information, like from different key substring hypotheses, different selection functions, and different processing rounds. It allows to exploit redundancy in the expanded key of multi-round cryptographic algorithms for soft-decision estimation error correction. Computational effort is reduced by using conditional independencies of variables, and scalable with sequential decoding methods.

The example implementation using the M-algorithm includes standard differential analysis as one special case (for $M_E = M_R = 1$ if subkey hypothesis length is one bit, and using only one round) and has strictly better estimation accuracy otherwise. For longer subkey hypothesis length the proposed algorithm is more accurate due to probabilistic combing of hypotheses likelihoods. It further performs better if multiple rounds are considered, because intermediate errors of early quantization are avoided and can be corrected over the path metric. This higher accuracy comes at the price of increased computational complexity.

Several modifications and extensions of the presented methods are possible. They could be combined with different side channel analysis algorithms other than single-bit differential analysis. Other sequential decoding implementations like the stack algorithm or T-algorithm are applicable as well. It is further possible to terminate early, i.e. not to follow all processing rounds (some proposed Turbo Decoder implementations for example use a stopping criterion based on LLR magnitudes). The presented example implementation also did not yet enforce structural constraints like the key expansion algorithm. Pruning the code tree can be improved by considering only valid expanded key sequences.

With sequential decoding, the side channel analysis problem becomes a 3-dimensional trade-off between measurement SNR, number of required traces and computational complexity. It becomes possible for example to use cheaper measurement equipment (worse SNR) by computing a bit more, or to a achieve a new minimum number of required traces by investing high computational effort. Future work may consist of further characterizing this trade-off by comparing algorithm modifications and identifying Pareto-efficient ones.